\documentclass[nofootinbib,prd,twocolumn,showpacs,showkeys,preprintnumbers]{revtex4-1}
\usepackage{hyperref,amssymb,amsmath,mathrsfs,bm,graphicx}
\begin{document}
\title {Complexity factors for axially symmetric static sources }
\author{L. Herrera}
\email{lherrera@usal.es}
\affiliation{Instituto Universitario de F\'isica
Fundamental y Matem\'aticas, Universidad de Salamanca, Salamanca 37007, Spain}
\author{A. Di Prisco}
\email{alicia.diprisco@ciens.ucv.ve}
\affiliation{Escuela de F\'\i sica, Facultad de Ciencias, Universidad Central de Venezuela, Caracas 1050, Venezuela}
\author{J. Ospino}
\email{j.ospino@usal.es}
\affiliation{Departamento de Matem\'atica Aplicada and Instituto Universitario de F\'isica
Fundamental y Matematicas, Universidad de Salamanca, Salamanca 37007, Spain}

\date{\today}
\begin{abstract}
A previously found definition of   complexity for  spherically symmetric  fluid distributions \cite{css}, is extended to axially symmetric static sources. In this case there are three different complexity factors, defined in terms of three structure scalars obtained from the orthogonal splitting of the Riemann tensor. All these three factors vanish, for what we consider the simplest fluid distribution, i.e a fluid  spheroid with isotropic pressure and homogeneous energy density.  However, as in the spherically symmetric case, they can also vanish for a variety of  configurations, provided the energy density inhomogeneity terms cancel the pressure anisotropic ones in the expressions for the complexity factors. Some exact analytical solutions of this type are found and analyzed. At the light of the obtained results, some conclusions  about the correlation (the lack of it)  between symmetry and complexity,  are put forward.
\end{abstract}
\date{\today}
\maketitle

\section{Introduction}
In recent  papers  a new definition of complexity, in the context of general relativity,  has been proposed for spherically symmetric fluid distributions, in the static case \cite{css} and the dynamical case \cite{csd}. Applications of this concept   to other theories of gravity have been proposed in  \cite{ot1, ot2}, while  the charged case  has been considered in \cite{csd} and \cite{ch}. Also, applications   for some particular cases of cylindrically symmetric fluid distributions, may be found in \cite{cil}. It is our goal in this work to extend this definition of complexity to the most general axially symmetric static fluid distributions.

The motivation for such an endeavour is based on the fact that while it is true that observational evidence seems to suggest  that deviations from spherical symmetry in compact
self-gravitating objects (white dwarfs, neutron stars), are likely to be incidental rather than basic features of these systems (putting aside the evident fact that astrophysical objects are generally endowed with angular momentum), it also true that there is a  bifurcation between any finite perturbation of Schwarzschild spacetime and any Weyl solution, even when the latter is characterized by parameters arbitrarily close to those corresponding to spherical symmetry (see \cite{i2}-\cite{in} and references therein for a discussion on this point). This fact in turn is related to the well known result  that the only regular static and asymptotically flat vacuum spacetime possessing a regular horizon is the Schwarzchild solution \cite{israel}, while all the others Weyl exterior solutions \cite{weyl2}-\cite{weyln} exhibit singularities in the  curvature invariants (as the boundary of the source approaches the horizon).

Sources of different Weyl spacetimes have already been considered by several authors in the past (see for example  \cite{1}-\cite{sn1} and references therein).

More recently,  a renewed interest on this kind of solutions have aroused, particularly in relation to the deviations of spherical symmetry produced by  different physical phenomena such as magnetic fields (see for example \cite{s3}-\cite{s8} and references therein).

In the spherically symmetric case, the complexity factor is a scalar variable intended to measure the degree of complexity of the fluid distribution. For reasons explained in  detail in \cite{css, csd}, such scalar function may be identified as one of the scalar functions (structure scalars)  which appears in the orthogonal splitting of the Riemann tensor  \cite{1cil}.  More specifically, it is related to one of the scalar functions appearing in the splitting of the electric part of the Riemann tensor (see also \cite{2cil, 3cil, 4cil, 5cil, scc1, scc2, csp, hu, mbi, ku, ah}  and references therein for further discusion on the structure scalar).

In the axially symmetric case the situation is much more complicated, and the number of structure scalars much larger than in the spherically symmetric case. Nevertheless the general criterium  to define the variable(s) measuring the complexity of the fluid distribution will be the same, namely: we start by asking ourselves which is the simplest fluid configuration.  As in the spherically symmetric case we shall assume that such a configuration corresponds to the incompressible (constant energy density), isotropic (in the pressure) spheroid. From this simple assumption, we shall see that as the obvious candidates to measure the degree of complexity of the fluid distribution, appear three of the eight structure scalars corresponding to the axially symmetric static fluid distribution. Explicit forms of these structure scalars as well as some useful differential  equations relating the inhomogeneities of the energy density to some of the structure scalars were already found in \cite{as}.

As in the spherically symmetric case, the vanishing of the three complexity factors corresponds not only to the incompressible, isotropic  spheroid, but also to a large family of solutions where the density inhomogeneity terms cancel the pressure anisotropic terms in the equations relating these to the complexity factors. Some of these solutions will be exhibited.

Our paper is organized as follows: In the next section we shall review the general framework developped in \cite{as} to describe  the most general non--vacuum,  axially symmetric static  spacetime. This include relevant variables and equations. Next we define the complexity factors for our system and discuss about their general properties. The two families of solutions found are described in sections IV.  A summary of the obtained results as well as a list of some  unsolved issues are presented in section V. Finally three  Appendices with some useful equations  are included.
\section{The general framework}
As we mentioned in the Introduction, a general framework for describing axially symmetric static sources was deployed in \cite{as}. Here we shall resort (with slight changes in the notation) to such a formalism. However in order to render this manuscript self--consistent  we shall provide in this section a brief resume  of  the approach to be used. The reader may find all the details in \cite{as}.
\subsection{The metric and the source}
We shall consider  static and axially symmetric sources. For such a system the line element may be written in ``Weyl spherical coordinates'' as:
\begin{equation}
ds^2=-A^2 dt^2 + B^2 \left(dr^2 +r^2d\theta^2\right)+D^2d\phi^2,
\label{1b}
\end{equation}
where the coordinates $t$ and $\phi$ are adapted to the two Killing vectors admitted by our line element, and therefore the metric functions depend only on $r$ and $\theta$.

For the sake of generality we shall not assume here the Weyl gauge.  In   the vacuum case this gauge  can be used without loss of generality, and it allows for  the reduction of  the line element so that only two independent metric functions appear. However, in the interior  this is not possible in general, though obviously one may assume it as an additional restriction, which amounts to satisfy $R^3_3+R^0_0=0$, where $R^\alpha_\beta$ denotes the Ricci tensor.

Let us now provide a full description of the source. In order to give physical significance to the components of the energy momentum tensor, we shall  apply the Bondi approach \cite{Bo}, which consists in defining the physical variables in a purely locally
Minkowski frame ($\tau, x, y, z$) (hereafter referred to as l.M.f.) where the first derivatives of the metric vanish (locally), or, equivalently,  consider a tetrad
field attached to such l.M.f.

For the system under consideration, the most general  energy--momentum tensor in such locally defined coordinate system is given by:

\begin{equation}
\widehat{T}_{\alpha\beta}= \left(\begin{array}{cccc}\mu    &  0  &   0     &   0    \\0 &  P_{xx}    &   P_{xy}     &   0    \\0       &   P_{yx} & P_{yy}  &   0    \\0       &   0       &   0     &   P_{zz}\end{array} \right) \label{3},
\end{equation}
\\
where $\mu, P_{xy}, P_{xx}, P_{yy}, P_{zz}$ denote the energy density and different stresses, respectively, as measured by our locally defined Minkowskian observer.

Also observe that  $P_{xy}= P_{yx} $ and, in general  $ P_{xx}  \neq  P_{yy}  \neq P_{zz}$.

Then transforming back to our coordinates, we obtain the components of the energy momentum tensor in terms of the physical variables as defined in the l.M.f.
\begin{eqnarray}
{T}_{\alpha\beta}&=& (\mu+P_{zz}) V_\alpha V_\beta+P_{zz} g _{\alpha \beta} +(P_{xx}-P_{zz}) K_\alpha  K_\beta\nonumber \\ &+& (P_{yy}-P_{zz}) L_\alpha L_\beta +2P_{xy} K_{(\alpha}  L_{\beta)},
\label{6}
\end{eqnarray}
where
\begin{eqnarray}
 V_\alpha=(-A,0,0,0);\quad  K_\alpha=(0,B,0,0);\nonumber \\
   L_\alpha=(0,0,Br,0); \quad S_{\alpha}=(0, 0, 0, D),
\label{7}
\end{eqnarray}
where we are considering observers at rest with respect to the fluid distribution.

Alternatively we may write the energy momentum tensor in the ``canonical'' form:
\begin{eqnarray}
{T}_{\alpha\beta}&=& (\mu+P) V_\alpha V_\beta+P g _{\alpha \beta} +\Pi_{\alpha \beta},
\label{6bis}
\end{eqnarray}
with
\begin{eqnarray}
\Pi_{\alpha \beta}&=&(P_{xx}-P_{zz})\left(K_\alpha K_\beta-\frac{h_{\alpha \beta}}{3}\right)\nonumber \\&+&(P_{yy}-P_{zz})\left(L_\alpha L_\beta-\frac{h_{\alpha \beta}}{3}\right)+2P_{xy}K_{(\alpha}L_{\beta)}
\label{6bb},
\end{eqnarray}
and
\begin{equation}
P=\frac{P_{xx}+P_{yy}+P_{zz}}{3}, \quad h_{\mu \nu}=g_{\mu \nu}+V_\nu V_\mu.
\label{7Pb}
\end{equation}
and:
\begin{widetext}
\begin{eqnarray}
 \Pi_{\alpha \beta}=\frac{1}{3}(2\Pi_I+\Pi_{II})\left( K_\alpha  K_\beta -\frac{ h_{\alpha
\beta}}{3}\right)+\frac{1}{3}(2\Pi _{II}+ \Pi_I)\left( L_\alpha   L_\beta -\frac{ h_{\alpha
\beta}}{3}\right)+ \Pi _{KL}\left( K_\alpha
 L_\beta+ K_\beta
 L_\alpha\right) \label{6bb},
\end{eqnarray}
\end{widetext}
with
\begin{eqnarray}
 \Pi _{KL}= K^\alpha  L^\beta T_{\alpha \beta},
 \label{7P}
\end{eqnarray}

\begin{equation}
 \Pi_I=\left(2  K^\alpha   K^\beta -   L^\alpha   L^\beta-  S^\alpha  S^\beta \right)  T_{\alpha \beta},
\label{2n}
\end{equation}
\begin{equation}
 \Pi_{II}=\left(2  L^\alpha  L^\beta - K^\alpha   K^\beta- S^\alpha   S^\beta \right) T_{\alpha \beta}.
\label{2nbis}
\end{equation}

The relationships between the above scalars and the variables $P_{xy}, P_{xx}, P_{yy}, P_{zz}$ are, (besides (\ref{7Pb})), as follows:
\begin{equation}
\Pi_2\equiv \frac{1}{3}(2\Pi_I+\Pi_{II})=P_{xx}- P_{zz},
\label{3nbis}
\end{equation}
\begin{equation}
\Pi_3\equiv \frac{1}{3}(2\Pi_{II}+\Pi_{I})=P_{yy}- P_{zz},
\label{4nbis}
\end{equation}
\begin{equation}
\Pi_{KL}=P_{xy}.
\label{5nbis}
\end{equation}
or, inversely:
\begin{equation}
 P_{zz}=P-\frac{1}{3}(\Pi_2+\Pi_3),
 \label{6nbis}
\end{equation}
\begin{equation}
 P_{xx}=P+\frac{1}{3}(2\Pi_2-\Pi_3),
\label{7nbis}
\end{equation}
\begin{equation}
P_{yy}=P+\frac{1}{3}(2\Pi_3-\Pi_2).
\label{8nbis}
\end{equation}

The explicit form of the Einstein equations as well as the conservation equations, for the line element (\ref{1b}) and the energy--momentum tensor (\ref{6}), are given in the Appendix A and B respectively.
\subsection{The structure scalars}
The structure scalars for our problem were calculated in \cite{as}.  For their definition we need  first to obtain the electric part of the Weyl tensor (the magnetic part vanishes identically), whose components  can be obtained directly from its definition,
\begin{equation}
E_{\mu\nu}=C_{\mu\alpha\nu\beta}\,V^\alpha\, V^\beta,\label{8}
\end{equation}
where $C_{\mu\alpha\nu\beta}$ denotes the Weyl tensor. These are exhibited in the Appendix B.

Equivalently, the electric part of the Weyl tensor may also be written as:
\begin{widetext}
\begin{eqnarray}
E_{\alpha \beta}=\mathcal{E}_1\left(K_\alpha L_\beta+L_\alpha K_\beta\right)
+\mathcal{E}_2\left(K_\alpha K_\beta-\frac{1}{3}h_{\alpha \beta}\right)+\mathcal{E}_3\left(L_\alpha L_\beta-\frac{1}{3}h_{\alpha \beta}\right), \label{13}
\end{eqnarray}
\end{widetext}
where explicit expressions for the three scalars $\mathcal{E}_1$, $\mathcal{E}_2$, $\mathcal{E}_3$ are given in the Appendix.

Next, let us calculate the electric part of the Riemann tensor (the magnetic part vanishes identically), which is defined by
\begin{equation}
Y^\rho_\beta=V^\alpha V^\mu R^\rho_{\alpha \beta \mu}.
\label{29}
\end{equation}

After some lengthy calculations we find;

\begin{eqnarray}
Y_{\alpha \beta}&=&Y_{TF_1}\left(K_\alpha L_\beta+K_\beta L_\alpha\right)
+Y_{TF_2}\left(K_\alpha K_\beta-\frac{1}{3}h_{\alpha \beta}\right)\nonumber \\
&+&Y_{TF_3}\left(L_\alpha L_\beta-\frac{1}{3}h_{\alpha \beta}\right)+\frac{1}{3} Y_T h_{\alpha \beta},
\label{30}
\end{eqnarray}
where
\begin{equation}
Y_T=4\pi(\mu+3P),
\label{31}
\end{equation}

\begin{equation}
Y_{TF_1}=\mathcal{E}_1-4\pi \Pi_{KL},
\label{32}
\end{equation}
\begin{equation}
Y_{TF_2}=\mathcal{E}_2-4\pi \Pi_2,
\label{33}
\end{equation}

\begin{equation}
Y_{TF_3}=\mathcal{E}_3-4\pi \Pi_3.
\label{34}
\end{equation}
Finally, we shall find the tensor associated with the double dual of Riemann tensor, defined as:
\begin{equation}
X_{\alpha \beta}=^*R^{*}_{\alpha \gamma \beta \delta}V^\gamma
V^\delta=\frac{1}{2}\eta_{\alpha\gamma}^{\quad \epsilon
\rho}R^{*}_{\epsilon \rho\beta\delta}V^\gamma V^\delta,
\label{35}
\end{equation}
with $R^*_{\alpha \beta \gamma \delta}=\frac{1}{2}\eta
_{\epsilon \rho \gamma \delta} R_{\alpha \beta}^{\quad \epsilon
\rho}$,
where $\eta_{\epsilon \rho \gamma \delta}$ denotes the permutation symbol.

Thus, we find
\begin{eqnarray}
X_{\alpha \beta}&=&X_{TF_1}\left(K_\alpha L_\beta+K_\beta L_\alpha\right)+
X_{TF_2}\left(K_\alpha K_\beta-\frac{1}{3}h_{\alpha \beta}\right)\nonumber \\
&+&X_{TF_3}\left(L_\alpha L_\beta-\frac{1}{3}h_{\alpha \beta}\right)+\frac{1}{3}X_T h_{\alpha \beta},
\label{36}
\end{eqnarray}
where
\begin{equation}
X_T=8\pi \mu,
\label{37}
\end{equation}

\begin{equation}
X_{TF_1}=-(\mathcal{E}_1+4\pi \Pi_{KL}),
\label{38}
\end{equation}
\begin{equation}
X_{TF_2}=-\left(\mathcal{E}_2+4\pi \Pi_2\right),
\label{39}
\end{equation}

\begin{equation}
X_{TF_3}=-\left(\mathcal{E}_3+4\pi \Pi_3\right).
\label{40}
\end{equation}

The  scalars $Y_T$, $Y_{TF1}$, $Y_{TF2}$,$ Y_{TF3}$, $X_T$, $X_{TF1}$, $X_{TF2}$, $X_{TF3}$, are the structure scalars for our problem.

\subsection{Some differential equations for the structure scalars }
Two differential equations which relate the spatial derivatives of the  physical variables and the Weyl tensor  may be obtained using Bianchi identities, they have been found before for the spherically symmetric and the cylindrically symmetric cases (see \cite{1cil}, \cite{scc1} and references therein). For our case they  have been calculated in \cite{as}:
\begin{widetext}
\begin{eqnarray}
\frac{{\cal E}_{1\theta}}{r}&+&\frac{1}{3}(2{\cal E}_2-{\cal E}_3)^\prime+\frac{{\cal E}_1}{r}\left(\frac{2B_\theta}{B}+
\frac{D_\theta}{D}\right)+{\cal E}_2\left(\frac{B^\prime}{B}+\frac{D^\prime}{D}+\frac{1}{r}\right) -{\cal E}_3 \left(\frac {B^\prime}{B}+\frac{1}{r}\right)=
\frac{4\pi}{3}\left(2\mu+3P\right)^\prime\\ \nonumber
&+&4\pi \left[\mu+P+\frac{1}{3}(2\Pi_2-\Pi_3)\right]\frac{A^\prime}{A}+4\pi \Pi_{KL}\frac{A_\theta}{Ar},
\label{55}
\end{eqnarray}
\end{widetext}
\begin{widetext}
\begin{eqnarray}
{\cal E}^{\prime}_1&+&\frac{1}{3r}(2{\cal E}_3-{\cal E}_2)_\theta+{\cal E}_1\left(\frac{2B^\prime}{B}+
\frac{D^\prime}{D}+\frac{2}{r}\right)-\frac{{\cal E}_2 B_\theta}{Br}+\frac{{\cal E}_3}{r} \left(\frac{B_\theta}{B}+\frac{D_\theta}{D}\right)=
\frac{4\pi}{3r}\left(2\mu+3P \right)_\theta\\ \nonumber &+&4\pi\left[\mu+P+\frac{1}{3}(2\Pi_{3}-\Pi_2)\frac{A_\theta}{Ar}\right]
+4\pi \Pi_{KL}\frac{A^\prime}{A},
\label{56}
\end{eqnarray}
\end{widetext}
which, using (\ref{31})-(\ref{34}) and (\ref{37})-{\ref{40}), may be written in terms of structure scalars:
\begin{widetext}
\begin{eqnarray}
\frac{8 \pi \mu^\prime}{3}&=&\frac{1}{r}\left[Y_{TF1\theta}+8\pi \Pi_{KL\theta} +(Y_{TF1}+8\pi \Pi_{KL})(\ln {B^2 D)}_{\theta}
\right]+\left[\frac{2}{3}(Y_{TF2}^\prime +8\pi \Pi_2^\prime)+(Y_{TF2}+8\pi \Pi_2)(\ln{BDr})^{\prime}\right]\nonumber \\
&-&\left[\frac{1}{3}(Y_{TF3}^\prime+8\pi\Pi_3^\prime)+(Y_{TF3}+8\pi \Pi_3)(\ln{Br})^{\prime}\right],
\label{57}
\end{eqnarray}
\end{widetext}
\begin{widetext}
\begin{eqnarray}
\frac{8 \pi \mu_{\theta}}{3r}&=&-\frac{1}{r}\left[\frac{1}{3}(Y_{TF2\theta}+8\pi \Pi_{2\theta})+(Y_{TF2}+8\pi \Pi_2)(\ln B)_{\theta}\right]+\frac{1}{r}
\left[\frac{2}{3}(Y_{TF3\theta}+8\pi \Pi_{3\theta})+(Y_{TF3}+8\pi \Pi_3)(\ln{ BD})_{\theta}\right]\nonumber \\&+&
\left[Y_{TF1}^\prime+8\pi \Pi_{KL}^\prime+(Y_{TF1}+8\pi \Pi_{KL})(\ln{B^2 D r^2})^{\prime}\right],
\label{58}
\end{eqnarray}
\end{widetext}
where prime  and subscript $\theta$ denote derivatives with respect to $r$  and $\theta$ respectively.
\section{The complexity factors}
We have now available all the elements necessary to define the complexity factors for  the fluid distribution under consideration. For doing so we have first to establish what we consider is the simplest possible fluid (or at least one of them). From elementary considerations, as we did in \cite{css, csd}, we assume that the incompressible (constant energy density) fluid with isotropic pressure is one of the simplest fluid distributions. Now, in \cite{as} it has been shown that  the necessary and sufficient conditions for the vanishing of the (invariantly defined) spatial derivatives of  the energy density  are $X_{TF1}=X_{TF2}=X_{TF3}=0$. In other words
\begin{equation}
X_{TF1}=X_{TF2}=X_{TF3}=0\Leftrightarrow  \mu^\prime=\mu_\theta=0.
\label{70n}\end{equation}

Therefore the homogeneous energy--density condition implies $X_{TF1}=X_{TF2}=X_{TF3}=0$, which in turn produces

\begin{equation}
Y_{TF1}=-8\pi \Pi_{KL};\quad Y_{TF2}=-8\pi \Pi_2;\quad Y_{TF3}=-8\pi \Pi_3.
\label{71n}\end{equation}

Obviously, the isotropic pressure condition would imply $Y_{TF1}= Y_{TF2}= Y_{TF3}=0$.

Thus from the above considerations, and following the rationale exposed in the spherically symmetric case, we shall identify the three structure scalars $Y_{TF}$ (more precisely, their absolute values) as the complexity factors. They vanish for the the incompressible (constant energy density) fluid with isotropic pressure, but may also vanish for inhomogeneous, anisotropic fluids, provided these two factors combine in such a way that they cancel the three complexity factors.

We shall next find some explicit analytical solutions.

\section{Solutions satisfying the vanishing complexity factors condition}
As was mentioned above, the fluid distribution with homogeneous energy density and isotropic pressure satisfies the vanishing complexity factors condition, but it is not the only one. These conditions may also be satisfied if the terms describing the energy density inhomogeneity cancel the anisotropic terms in (\ref{57}), (\ref{58}).

In this section we shall present some solutions of this kind. It should be kept in mind that our purpose here is not to present  solutions representing specific physically meaningful compact object, but just  to illustrate the way by means of which such models might be obtained.

\subsection{The incompressible, isotropic  spheroid}

As we have already seen, the incompressible isotropic spheroid represents a fluid distribution for which the three complexity factors vanish. This solution was obtained and analyzed in \cite{as}. Here we just reproduce it without details.  Thus from (\ref{70n}), (\ref{38})--(\ref{40}) and $P_{xx}=P_{yy}=P_{zz}=P$, $P_{xy}=0$, $\mu=\mu_0=constant$, it is evident that such a solution is also conformally flat.

For simplicity we shall assume the boundary surface $\Sigma$ to be defined by the equation:
\begin{equation}
r=r_1=constant.
\label{50ns}
\end{equation}

From  the above and  (\ref{25}) and (\ref{26}) it follows that
\begin{equation}
P\stackrel{\Sigma}{=}0,
\label{51ns}
\end{equation}
where $\stackrel{\Sigma}{=}$ means that both sides of the equation
are evaluated on $\Sigma$

Under the conditions above (\ref{22}) and (\ref{23}) can be integrated to obtain:
\begin{equation}
P+\mu_0=\frac{\zeta}{A},
\label{52}
\end{equation}
and
\begin{equation}
P+\mu_0=\frac{\xi(r)}{A},
\label{53}
\end{equation}
where $\xi$ is  an arbitrary function of its argument.
Using boundary conditions (\ref{51ns}) in (\ref{52}) (\ref{53}) it follows that:
\begin{equation}
 A(r_1,\theta)=const.=\frac{\alpha}{\mu_0}, \qquad \zeta=constant.
\label{54}
\end{equation}
Finally, the metric of incompressible conformally flat isotropic
fluids can be written as follows.

\begin{widetext}
 \begin{equation}
 ds^2=\frac{1}{(\gamma r^2+\delta +b r\cos \theta)^2}\left [-(\alpha r^2+\beta+a r\cos\theta)^2dt^2+dr^2+r^2d\theta ^2+r^2\sin ^2 \theta d\phi ^2\right ]. \label{cfifm}
\end{equation}
\end{widetext}

Next,  the physical variables can be easily calculated.  Thus, using (\ref{cfifm}) into (\ref{24}) the energy density reads:

 \begin{equation}
8\pi \mu = 12 \gamma \delta-3b^2.
\label{den1}
\end{equation}

To obtain  the pressure we shall use (\ref{52}) and (\ref{54}), which produce
\begin{equation}
8\pi P =(3b^2-12\gamma\delta )\left [1-\frac{\alpha
r_1^2+\beta}{\gamma r_1^2+\delta} \frac{\gamma
r^2+\delta+br\cos\theta}{\alpha r^2+\beta +a
r\cos\theta}\right ],\label{epf}
\end{equation}
where  $b, \gamma, \delta$ are constants, and 

\begin{equation}
\zeta=\mu_0\frac{\alpha r_1^2+\beta}{\gamma
r_1^2+\delta},\quad  a=\frac{\alpha r_1^2+\beta}{\gamma
r_1^2+\delta}b,
\label{jcn}
\end{equation}
in order to satisfy  the junction condition  (\ref{51ns}).

It is important to stress the fact that this solution cannot be matched to any Weyl exterior, except in the spherically symmetric case, even though it  has a surface of vanishing pressure (see \cite{csd} for details). This result is in agreement with  theorems indicating that static, perfect fluid (isotropic in pressure) sources are spherical (see \cite{prueba} and references therein).

\subsection{Anisotropic inhomogeneous spheroids}

\noindent  Although the inhomogeneous anisotropic spheroids exhibited in \cite{as} do not satisfy the vanishing complexity factors conditions, solutions with vanishing complexity factors but inhomogeneous energy density and anisotropic pressure do exist, as we shall show in this subsection.

The metric variables for the solution are:
\begin{eqnarray}
  A(r,\theta) &=& \frac{a_1 r \sin \theta}{b_1 r^2+b_2},\label{ns2a} \\
  B(r,\theta)&=& \frac{1}{b_1 r^2+b_2} \\
  D(r,\theta) &=& \frac{b_1 r^2-b_2}{b_1 r^2+b_2} F\left(\frac{r \cos \theta}{b_1 r^2-b_2}\right). \label{ns1a}
\end{eqnarray}
It is a simple matter to check that conditions (\ref{YTF1})--(\ref{YTF3}) are satisfied for (\ref{ns2a})--(\ref{ns1a}).

Next, using the Einstein equations (\ref{24})--(\ref{26}), the metric above produces the following expressions for the physical variables.
\begin{widetext}
\begin{eqnarray}
  8\pi \mu &=& 12b_1 b_2-\frac{(b_1 r^2+b_2)^2}{(b_1 r^2-b_2)^2}(\frac{4b_1 b_2  r^2 cos^2\theta}{(b_1 r^2-b_2)^2}+1)\frac{F_{zz}}{F}, \\
   8\pi P&=& -12b_1 b_2+\frac{(b_1 r^2+b_2)^2}{3(b_1 r^2-b_2)^2}(\frac{4 b_1 b_2  r^2 cos^2\theta}{(b_1 r^2-b_2)^2}+1)\frac{F_{zz}}{F} ,\\
  8\pi \Pi_2\equiv  8\pi (P_{xx}-P_{zz}) &=& \frac{F_{zz}}{4F}\frac{(b_1 r^2+b_2)^2}{(b_1 r^2-b_2)^2} \sin^2 \theta,\\
  8\pi \Pi_3\equiv  8\pi (P_{yy}-P_{zz}) &=& \frac{F_{zz}}{4F}\frac{(b_1 r^2+b_2)^4 cos^2\theta}{(b_1 r^2-b_2)^4},\\
   8\pi \Pi_{KL}\equiv 8\pi P_{xy}&=&- \frac{F_{zz}}{2F}\frac{r(b_1 r^2+b_2)^3}{(b_1 r^2-b_2)^3} \sin 2\theta,
\end{eqnarray}
\end{widetext}

\noindent with
\begin{equation}
F(z)\equiv F\left(\frac{r \cos\theta}{b_1 r^2-b_2}\right),
\end{equation}
and where $a_1, b_1, b_2$ are constants. For a range of values of these parameters, the physical behaviour of physical variables is acceptable and the metric may be matched smoothly on the  boundary surface to a Weyl solution. However, our only purpose in this section is to illustrate the existence of solutions admitting the vanishing complexity factors  condition, and not to model specific astrophysical objects.

\section{conclusions}
We have extended  a  previously proposed definition of complexity for spherically symmetric fluid distributions, to the axially symmetric static case. We have considered the most general fluid distribution compatible with this latter symmetry. Unlike the spherically symmetric case, the complexity is now defined in terms of three scalar functions (complexity factors). This fact opens  the possibility  to establish a more elaborated hierarchy of models, which runs from the simplest case (the vanishing of the three complexity factors) through semi--simple (semi--complex) models with only one or two vanishing complexity factors, until the more complex (the less simple) models with all the three complexity factors different from zero. Also, it is worth noticing that the three scalars $Y_{TF,1,2,3}$ may be positive or negative (if they are non--vanishing), depending on the interplay between energy density inhomogeneity and pressure anisotropy. Accordingly it is evident from purely physical considerations that we have to choose the absolute values of these scalars as the measure of the complexity of any fluid distribution.

As it happens in the spherically symmetric case, there are more than one model compatible with the vanishing of all the complexity factors. It remains as a pending task to find out what all these models have in common (besides the fact that the complexity factors vanish). In other words, what are the physical consequences derived from the vanishing of  the complexity factors?  In the spherically symmetric case the consequence derived from the vanishing of the complexity factor is very simple: the distribution of the Tolman (active gravitational) mass is the same for  all these configurations. We don't know if something similar appears in the axially symmetric case.

In relation to the comment above, we would like to stress one point which deserves to be explored in some detail: we refer to the study of the possible relationship between complexity (as defined here) and the stability of the fluid distribution. Such  relationship is apparent in the spherically symmetric case through the influence of the complexity factor in the value of the active gravitational mass (Tolman) within the fluid distribution.

Finally we would like to call the attention to an issue which may be relevant in the discussion about the definition of complexity. We have in mind here the possible link between symmetry (expressed through the admittance of Killing vectors), and complexity. Indeed, even though, at purely intuitive level, one might expect these two concepts to be closely intertwined, the fact is that our results in this work as well as in \cite{css, csd} point in the opposite direction. 

In the spherically symmetric case, both in the static and  in the dynamic case, there are three Killing vectors which are compatible with a broad hierarchy in the degree of complexity. The situation analyzed in this manuscript reinforces further this picture, by admitting a wider hierarchy of complexity, for a lesser degree of symmetry. 

Furthermore, there is  an example that illustrates the lack of correlation between symmetry  and complexity, in a particularly sharp and forceful  way. Such an example is provided by the Szekeres spacetime \cite{1sz, 2sz}. These are time dependent metrics sourced by pure dust, which in general do not admit a single Killing vector \cite{3sz}. However, in spite of the absence of symmetry, the electric part of the Weyl tensor is defined through a single scalar function \cite{4sz}. Then, since there are no pressure terms, if we restrict ourselves  to the class of axially symmetric Szekeres metrics, we conclude that there is only one complexity factor (see eqs.(43-44) in \cite{5sz}), as in the spherically symmetric case, revealing thereby that its degree of complexity is low, in spite of the fact that there is only one Killing vector (in the axially symmetric case). 

Thus the qualification of ``quasispherical'' assigned by Szekeres himself to his solution appears to be  well justified, due to the similar degree of complexity of both spacetimes. In other words, the concept of complexity adopted here, seems to represent better  than its symmetry,  some  deeper aspects of the system.
\begin{acknowledgments}
This work was partially
supported by the Spanish Ministry of Science and Innovation (grant
FIS2010-15492). J.O. acknowledges financial support from the Spanish
Ministry of Science and Innovation (grant FIS2009-07238).
\end{acknowledgments}
\appendix
\section{The Einstein and the conservation equations}
For the line element (\ref{1b}) and the energy momentum (\ref{6}), the Einstein equations read:
\begin{widetext}
\begin{eqnarray}
8\pi\mu=-\frac{1}{B^2}\left\{\frac{B^{\prime \prime}}{B}+\frac{D^{\prime \prime}}{D}+\frac{1}{r}\left(\frac{B^\prime}{B} +\frac{D^\prime}{D}\right)-\left(\frac{B^\prime}{B}\right)^2+\frac{1}{r^2}\left[\frac{B_{\theta \theta}}{B}+\frac{D_{\theta \theta}}{D}-\left(\frac{B_\theta}{B}\right)^2\right] \right\},
\label{24}
\end{eqnarray}
\end{widetext}
\begin{widetext}
\begin{eqnarray}
8\pi P_{xx}=\frac{1}{B^2}\left[\frac{A^\prime B^\prime}{AB}+ \frac{A^\prime D^\prime}{AD}+\frac{B^\prime D^\prime}{BD}+\frac{1}{r}\left(\frac{A^\prime}{A}+\frac{D^\prime}{D}\right)+\frac{1}{r^2}\left(\frac{A_{\theta \theta}}{A}+\frac{D_{\theta \theta}}{D}-\frac{A_\theta B_\theta}{AB}+\frac{A_\theta D_\theta}{AD}-\frac{B_\theta D_\theta}{BD}\right)\right],
\label{25}
\end{eqnarray}
\end{widetext}
\begin{widetext}
\begin{eqnarray}
8\pi P_{yy}=\frac{1}{B^2}\left[\frac{A^{\prime \prime}}{A}+ \frac{D^{\prime \prime}}{D}-\frac{A^\prime B^\prime}{AB} +\frac{A^\prime D^\prime}{AD}-\frac{B^\prime D^\prime}{BD}+\frac{1}{r^2}\left(\frac{A_\theta B_\theta}{AB}+\frac{A_\theta D_\theta}{AD}+\frac{B_\theta D_\theta}{BD}\right)\right],
\label{27}
\end{eqnarray}
\end{widetext}
\begin{widetext}
\begin{eqnarray}
8\pi P_{zz}=\frac{1}{B^2}\left\{\frac{A^{\prime \prime}}{A}+ \frac{B^{\prime \prime}}{B}-\left(\frac{B^\prime}{B}\right)^2+\frac{1}{r}\left(\frac{A^\prime}{A} +\frac{B^\prime}{B}\right) +\frac{1}{r^2}\left[\frac{A_{\theta \theta}}{A}+\frac{B_{\theta \theta}}{B}-\left(\frac{B_\theta}{B}\right)^2\right]\right\},
\label{28}
\end{eqnarray}
\end{widetext}
\begin{widetext}
\begin{eqnarray}
8\pi P_{xy}=\frac{1}{B^2}\left\{  \frac{1}{r}\left[-\frac{A^{\prime}_\theta}{A} -\frac{D^{\prime}_\theta}{D} +\frac{B_\theta}{B}\left(\frac{A^\prime}{A}+\frac{D^\prime}{D}\right)+\frac{B^\prime}{B}\frac{A_\theta}{A}+\frac{B^\prime}{B}\frac{D_\theta}{D}\right]+\frac{1}{r^2} \left(\frac{A_\theta}{A}+\frac{D_\theta}{D}\right)\right\}.\label{26}
\end{eqnarray}
\end{widetext}

The nonvanishing components of the conservation equations $T^{\alpha  \beta}_{;\beta}=0$ yield: the trivial equation
\begin{equation}
\dot \mu=0,
\label{21}
\end{equation}
where the overdot denotes derivative with respect to $t$,
and the two hydrostatic equilibrium equations 
\begin{widetext}
\begin{eqnarray}
\left[P+\frac{1}{3}(2\Pi_2-\Pi_3)\right]^\prime &+&\frac{A^{\prime}}{A}\left[\mu+P+\frac{1}{3}(2\Pi_2-\Pi_3)\right]+\frac{B^{\prime}}{B}(\Pi_2-\Pi_3)+\frac{D^{\prime}}{D}\Pi_2 \nonumber \\  &+& \frac{1}{r}\left[\left(\frac{A_\theta}{A}+2\frac{B_\theta}{B}+\frac{D_\theta}{D}\right)\Pi_{KL}+\Pi_{Kl\theta}+\Pi_2-\Pi_3\right]=0,
\label{22}
\end{eqnarray}

\begin{eqnarray}
\left[P+\frac{1}{3}(2 \Pi_3-\Pi_2)\right]_\theta&+&\frac{A_{\theta}}{A}\left[\mu +P+\frac{1}{3}(2 \Pi_3-\Pi_2)\right]+\frac{B_{\theta}}{B}(\Pi_3-\Pi_2)\nonumber \\&+&\frac{D_{\theta}}{D}\Pi_3+r\left[\left(\frac{A^{\prime}}{A}+2\frac{B^{\prime}}{B}+\frac{D^{\prime}}{D}\right)\Pi_{KL}+\Pi^{\prime}_{KL}\right]+2\Pi_{KL}=0.
\label{23}
\end{eqnarray}
\end{widetext}
\section{Expression for the components of the electric Weyl tensor}
There are four  nonvanishing components as calculated from (\ref{8}), however they are not independent since they satisfy the relationship:
\begin{equation}
E_{11}+\frac{1}{r^2}E_{22}+\frac{B^2}{D^2}E_{33}=0,
\end{equation}
implying that the Weyl tensor may be expressed through three independent scalar functions $\mathcal{E}_1, \mathcal{E}_2, \mathcal{E}_3$.

These four components are
\begin{widetext}
\begin{eqnarray}
E_{11} &=& \frac{1}{6}\left[\frac{2A^{\prime \prime}}{A}-\frac{B^{\prime \prime}}{B}-\frac{D^{\prime \prime}}{D}-\frac{3A^{\prime} B^{\prime}}{AB}-\frac{A^{\prime} D^{\prime}}{AD}+\left(\frac{B^{\prime}}{B}\right)^2+\frac{3B^{\prime} D^{\prime}}{BD}+\frac{1}{r}\left(2\frac{D^\prime}{D}-\frac{B^\prime}{B}-\frac{A^\prime}{A}\right)\right]\nonumber \\
&+&\frac{1}{6r^2}\left[-\frac{A_{\theta \theta}}{A} -\frac{B_{\theta \theta}}{B} +\frac{2 D_{\theta \theta}}{D} +\frac{3A_\theta B_\theta}{AB} -\frac{A_\theta D_\theta}{AD}+\left(\frac{B_\theta}{B}\right)^2-\frac{3B_\theta D_\theta}{BD}\right],\label{9}
\end{eqnarray}

\begin{eqnarray}
E_{22} &=& -\frac{r^2}{6}\left[\frac{A^{\prime \prime}}{A}+\frac{B^{\prime \prime}}{B}-\frac{2D^{\prime \prime}}{D}-\frac{3A^{\prime} B^{\prime}}{AB}+\frac{A^{\prime} D^{\prime}}{AD}-\left(\frac{B^{\prime}}{B}\right)^2+\frac{3B^{\prime} D^{\prime}}{BD}+\frac{1}{r}\left(\frac{D^\prime}{D}+\frac{B^\prime}{B}-\frac{2A^\prime}{A}\right)\right]\nonumber \\
&-&\frac{1}{6}\left[-\frac{2A_{\theta \theta}}{A} +\frac{B_{\theta \theta}}{B} +\frac{ D_{\theta \theta}}{D} +\frac{3A_\theta B_\theta}{AB} +\frac{A_\theta D_\theta}{AD}-\left(\frac{B_\theta}{B}\right)^2-\frac{3B_\theta D_\theta}{BD}\right],\label{10}
\end{eqnarray}

\begin{eqnarray}
E_{33} &=& -\frac{D^2}{6B^2}\left[\frac{A^{\prime \prime}}{A}-\frac{2B^{\prime \prime}}{B}+\frac{D^{\prime \prime}}{D}-\frac{2A^{\prime} D^{\prime}}{AD}+2\left(\frac{B^{\prime}}{B}\right)^2+\frac{1}{r}\left(\frac{D^\prime}{D}-\frac{2B^\prime}{B}+\frac{A^\prime}{A}\right)\right]\nonumber \\
&-&\frac{D^2}{6B^2r^2}\left[\frac{A_{\theta \theta}}{A} -\frac{2B_{\theta \theta}}{B} +\frac{ D_{\theta \theta}}{D}  -\frac{2A_\theta D_\theta}{AD}+2\left(\frac{B_\theta}{B}\right)^2\right],\label{11}
\end{eqnarray}

\begin{eqnarray}
E_{12}  = \frac{1}{2} \left[\frac{A^{\prime}_\theta}{A} -\frac{D^{\prime}_\theta}{D}+\frac{B_\theta}{B}\frac{D^{\prime}}{D}-\frac{A^\prime B_{\theta}}{AB}-\frac{B^\prime A_{\theta}}{AB}+\frac{D_\theta}{D}\frac{B^{\prime}}{B}-\frac{1}{r}\left(\frac{A_\theta}{A}-
\frac{D_\theta}{D}\right)\right] \label{12}.
\end{eqnarray}
\end{widetext}

For the three scalars $\mathcal{E}_1, \mathcal{E}_2, \mathcal{E}_3$ we obtain:

\begin{widetext}
\begin{eqnarray}
\mathcal{E}_1= \frac{1}{2B^2} \left[\frac{1}{r}\left(\frac{A^{\prime}_\theta}{A} -\frac{D^{\prime}_\theta}{D}-
\frac{B_\theta}{B}\frac{A^{\prime}}{A}+\frac{D^{\prime}}{D}\frac{B_{\theta}}{B}-\frac{B^\prime}{B}\frac{A_\theta}{A}+
\frac{D_\theta}{D}\frac{B^\prime}{B}\right)+\frac{1}{r^2}\left(\frac{D_{\theta}}{D}-\frac{A_\theta}{A}\right)\right],\label{15}
\end{eqnarray}

\begin{eqnarray}
\mathcal{E}_2 & = &-\frac{1}{2B^2}\left[-\frac{A^{\prime \prime}}{A}+\frac{B^{\prime \prime}}{B}+
\frac{A^\prime B^\prime}{AB}+\frac{A^\prime D^\prime}{AD}-\left(\frac{B^\prime}{B}\right)^2-\frac{B^\prime D^\prime}{BD}+\frac{1}{r}\left(\frac{B^\prime}{B}
-\frac{D^\prime}{D}\right)\right]\nonumber \\ &-&\frac{1}{2B^2r^2}\left[\frac{B_{\theta \theta}}{B} -
\frac{D_{\theta \theta}}{D} -\frac{A_{\theta}B_{\theta}}{AB} +\frac{A_\theta D_\theta}{AD} -
\left(\frac{B_\theta}{B}\right)^2+\frac{B_\theta D_\theta}{BD}\right],\label{16}
\end{eqnarray}

\begin{eqnarray}
\mathcal{E}_3 & = &-\frac{1}{2B^2}\left[\frac{B^{\prime \prime}}{B}-\frac{D^{\prime \prime}}{D}-
\frac{A^\prime B^\prime}{AB}+\frac{A^\prime D^\prime}{AD}-\left(\frac{B^\prime}{B}\right)^2+\frac{B^\prime D^\prime}{BD}+\frac{1}{r}\left(\frac{B^\prime}{B}
-\frac{A^\prime}{A}\right)\right]\nonumber \\ &-&\frac{1}{2B^2r^2}\left[\frac{B_{\theta \theta}}{B} -\frac{A_{\theta \theta}}{A}
 +\frac{A_{\theta}B_{\theta}}{AB} +\frac{A_\theta D_\theta}{AD} -
\left(\frac{B_\theta}{B}\right)^2-\frac{B_\theta D_\theta}{BD}\right].\label{17}
\end{eqnarray}
\end{widetext}
Or, using Einstein equations we may also write:
\begin{widetext}
\begin{eqnarray}
\mathcal{E}_1 =\frac{E_{12}}{B^2r}=4\pi \Pi_{KL}+\frac{1}{B^2 r}\left[\frac{A^{\prime}_\theta}{A}-
\frac{A^{\prime} B_\theta}{AB}-\frac{A_\theta}{A}\left (\frac{B^{\prime}}{B}+\frac{1}{r}\right)\right],
\label{18}
\end{eqnarray}

\begin{eqnarray}
\mathcal{E}_2 &=&-\frac{2E_{33}}{D^2}-\frac{E_{22}}{B^2r^2}=
{4\pi} (\mu+3P+\Pi_2)
-\frac{A^\prime}{B^2A}\left(\frac{2D^{\prime}}{D}
+\frac{B^{\prime}}{B}+\frac{1}{r}\right)\nonumber \\
&+&\frac{A_\theta}{AB^2r^2} \left(\frac{B_\theta}{B}-\frac{2D_\theta}{D}\right)-\frac{1}{B^2r^2}\frac{A_{\theta \theta}}{A},
\label{19}
\end{eqnarray}

\begin{eqnarray}
\mathcal{E}_3 =-\frac{E_{33}}{D^2}+\frac{E_{22}}{B^2r^2}=4\pi \Pi_3
-\frac{A^\prime}{B^2A}\left(\frac{D^{\prime}}{D}
-\frac{B^{\prime}}{B}-\frac{1}{r}\right)
-\frac{A_\theta}{AB^2r^2} \left(\frac{D_\theta}{D}+\frac{B_\theta}{B}\right)+
\frac{1}{B^2r^2}\frac{A_{\theta \theta}}{A}.
\label{20}
\end{eqnarray}
\end{widetext}
\section{Vanishing complexity conditions}
\begin{widetext}
\begin{eqnarray}
Y_{TF_1}&=&\frac{1}{B^2 r}\left[\frac{A^{\prime}_\theta}{A}-
\frac{A^{\prime} B_\theta}{AB}-\frac{A_\theta}{A} \left(\frac{B^{\prime}}{B}+\frac{1}{r}\right)\right]=0,
\label{YTF1}
\end{eqnarray}

\begin{eqnarray}
Y_{TF_2} &=&\frac{A^{\prime\prime}}{B^2A}
-\frac{A^\prime}{B^2A}\left(\frac{D^{\prime}}{D}
+\frac{B^{\prime}}{B}\right)+\frac{A_\theta}{AB^2r^2} \left(\frac{B_\theta}{B}-\frac{D_\theta}{D}\right)=0,
\label{YTF2}
\end{eqnarray}

\begin{eqnarray}
Y_{TF_3} &=&
-\frac{A^\prime}{B^2A}\left(\frac{D^{\prime}}{D}
-\frac{B^{\prime}}{B}-\frac{1}{r}\right)
-\frac{A_\theta}{AB^2r^2} \left(\frac{D_\theta}{D}+\frac{B_\theta}{B}\right)+
\frac{1}{B^2r^2}\frac{A_{\theta \theta}}{A}=0.
\label{YTF3}
\end{eqnarray}
\end{widetext}

 
\end{document}